\documentclass[sigconf,screen]{acmart}
\renewcommand\footnotetextcopyrightpermission[1]{}
\acmConference{TREC 2026}{November 2026}{Gaithersburg, MD, USA} \acmYear{2026}
\acmISBN{} \acmDOI{}

\usepackage{booktabs}
\usepackage{multirow}
\usepackage{amsmath}
\usepackage{microtype}
\usepackage{seqsplit}
\usepackage{tabularx}
\usepackage{tikz}
\usepackage{xspace}
\usetikzlibrary{arrows.meta,calc}

\newcommand{\PromptTag}{\textsuperscript{\textsf{(P)}}\xspace}
\newcommand{\ModelTag}{\textsuperscript{\textsf{(L)}}\xspace}
\newcommand{\CodeTag}{\textsuperscript{\textsf{(C)}}\xspace}

\title{RMIT-ADM+S at TREC RAG 2026: Research Agents, Retriever Ablations, and
Open-Weight Replication}

\author{Oleg Zendel}
\affiliation{%
  \institution{RMIT University}
  \city{Melbourne}
  \country{Australia}
}
\author{Kun Ran}
\affiliation{%
  \institution{RMIT University}
  \city{Melbourne}
  \country{Australia}
}
\author{Chenglong Ma}
\affiliation{%
  \institution{RMIT University}
  \city{Melbourne}
  \country{Australia}
}
\author{Nguyen Khoi Vo}
\affiliation{%
  \institution{RMIT University}
  \city{Melbourne}
  \country{Australia}
}
\author{Shuai Wang}
\authornote{Now at keenable.ai}
\affiliation{%
  \institution{The University of Queensland}
  \city{Brisbane}
  \country{Australia}
}

\begin{document}

\begin{abstract}
  We describe the \texttt{RMIT-ADM+S} team's participation in the TREC 2026
  Retrieval-Augmented Generation (RAG) track using the ClimbMix corpus. We
  submitted seven research-agent runs using proprietary OpenAI and open-weight
  models, exploring different retrieval strategies, writer models, and agent
  workflows. Pre-submission ablations identified writer selection and
  adjacent-chunk expansion as the main factors affecting answer quality. In the
  official NIST evaluation, six of seven runs exceeded the track median across all
  three rubric/nugget collections. Our best-performing run, \texttt{brief-hybrid},
  achieved the highest all-item coverage among our submissions and tied the track
  maximum (0.892) for Batch-1 vital coverage. Human side-by-side preferences
  differed across evaluation batches: the open-weight \texttt{owa-glm5} led
  Batch~1 despite lower automated coverage, while \texttt{sol-research-first} led
  Batch~2. These findings highlight differences between automated coverage
  measures and human preferences, as well as the potential of lower-cost
  open-weight models for research-agent systems.
\end{abstract}

\keywords{retrieval-augmented generation, TREC RAG, agentic retrieval,
  Reciprocal Rank Fusion, open-weight models, trajectory fine-tuning}

\maketitle

\section{Introduction}
\label{sec:introduction}

We describe our participation in the TREC 2026 RAG track~\cite{trec2026rag},
using the \texttt{ClimbMix-400B-shuffle} corpus~\cite{diao2025climb}. The task
involves producing cited reports of up to 1,024 words for 119 test narratives.
We submitted seven research-agent runs using proprietary OpenAI and open-weight
models, exploring different retrieval strategies, writer models, and agent
workflows. These comparisons help assess whether more complex agent workflows
and proprietary models justify their respective computational and monetary
costs. This work complements our recent RAG competition systems, which explored
generation-assisted retrieval and reranking in LiveRAG and adaptive retrieval
with lightweight open-weight models in MMU-RAG~\cite{ran2025liverag,ran2026mmu}.

Our investigation focuses on three research questions: \textbf{RQ1:} Can
open-weight agents achieve comparable quality to proprietary models at lower
cost, and what explains any performance gaps? \textbf{RQ2:} Which model,
retrieval, and orchestration choices have the greatest impact on quality and
token cost? \textbf{RQ3:} Does multi-stage orchestration improve upon a simpler
single-loop agent or compensate for a weaker writer?

We evaluate these questions using pre-submission RAGDoll-based diagnostics
(Section~\ref{sec:results}) and official NIST evaluations covering 17 Batch-1
and 25 Batch-2 narratives (Section~\ref{sec:official}). Contrary to expectations
from our automated pre-submission evaluation, human preferences differed across
evaluation settings: the open-weight \texttt{owa-glm5} was significantly
preferred under holistic evaluation in Batch~1, while the proprietary
\texttt{sol-research-first} was preferred under nugget-guided evaluation in
Batch~2. This highlights how system rankings can differ not only between
automated and human evaluations, but also between human evaluations conducted
under different guidelines.

\section{System Overview}
\label{sec:system}

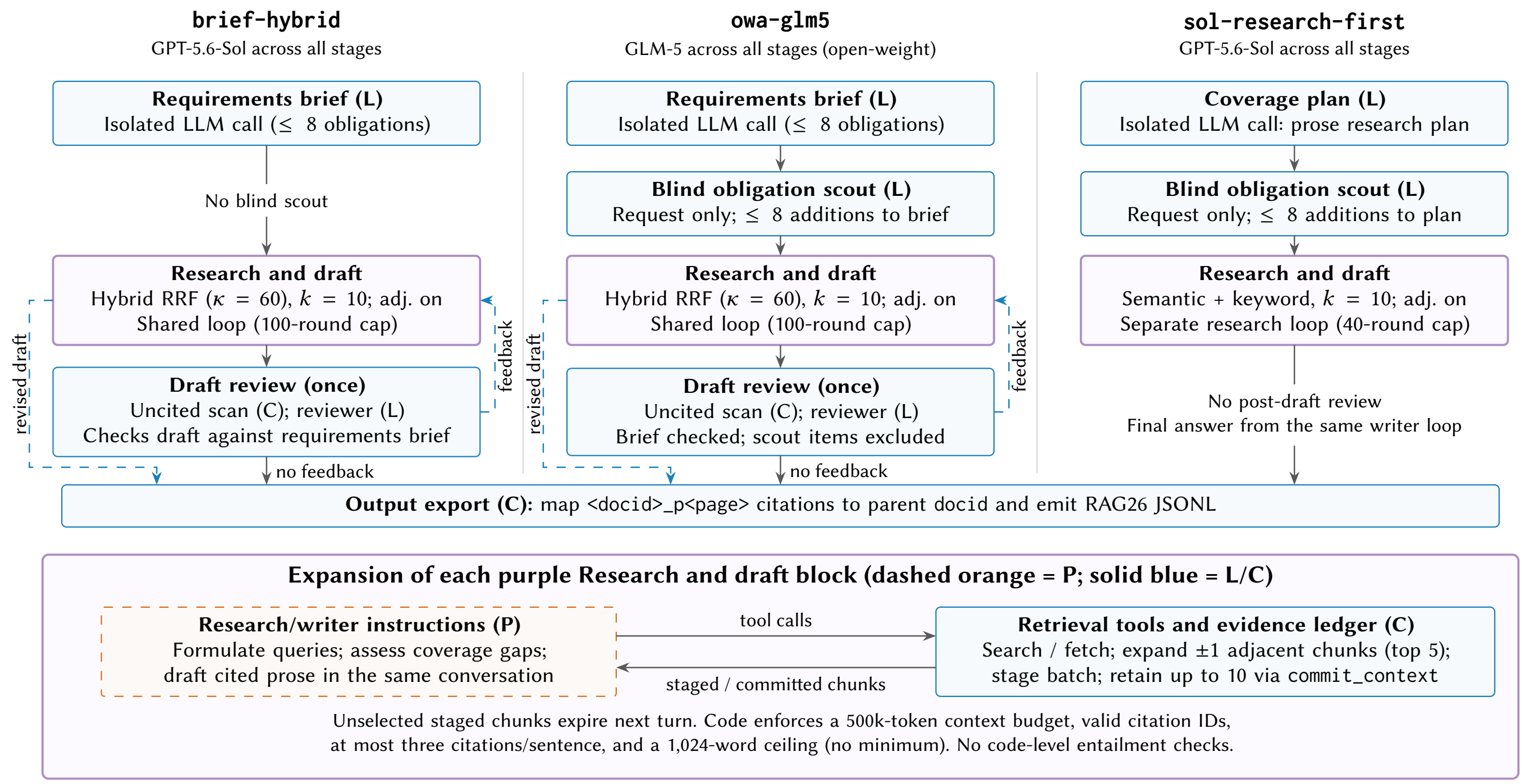
\begin{figure*}[t]
  \centering
  \renewcommand{\PromptTag}{\space(P)}
  \renewcommand{\ModelTag}{\space(L)}
  \renewcommand{\CodeTag}{\space(C)}
  \definecolor{figureblue}{RGB}{26,128,186}
  \definecolor{figureorange}{RGB}{217,110,26}
  \definecolor{mainblockcolor}{RGB}{215,180,243} 
  \resizebox{\textwidth}{!}{%
    \begin{tikzpicture}[x=1cm,y=1cm,
        stage/.style={draw=figureblue,fill=figureblue!4,line width=.4pt,
            rounded corners=1.5pt,align=center,text width=4.65cm,minimum height=.72cm,
            font=\sffamily\fontsize{7}{8.2}\selectfont,inner sep=3pt},
        research/.style={stage,draw=mainblockcolor!80!black,fill=mainblockcolor!7,line width=.7pt},
        prompt/.style={draw=figureorange,fill=figureorange!4,dashed,line width=.4pt,
            rounded corners=1pt,align=center,font=\sffamily\fontsize{7}{8.2}\selectfont},
        note/.style={align=center,font=\sffamily\fontsize{6.4}{7.8}\selectfont},
        heading/.style={align=center,font=\sffamily\bfseries\fontsize{8}{9.2}\selectfont},
        flow/.style={-{Stealth[length=1.5mm]},line width=.45pt,draw=black!65},
        conditional/.style={flow,dashed,draw=figureblue},
        every node/.style={outer sep=0pt}]

      \node[heading] at (2.55,0) {\texttt{brief-hybrid}};
      \node[heading] at (8.4,0) {\texttt{owa-glm5}};
      \node[heading] at (14.25,0) {\texttt{sol-research-first}};
      \node[note] at (2.55,-.33) {GPT-5.6-Sol across all stages};
      \node[note] at (8.4,-.33) {GLM-5 across all stages (open-weight)};
      \node[note] at (14.25,-.33) {GPT-5.6-Sol across all stages};
      \draw[black!15] (5.47,-.58) -- (5.47,-5.15);
      \draw[black!15] (11.32,-.58) -- (11.32,-5.15);

      \node[stage] (bbrief) at (2.55,-1.05)
      {\textbf{Requirements brief\ModelTag}\\Isolated LLM call ($\le 8$ obligations)};
      \node[stage] (obrief) at (8.4,-1.05)
      {\textbf{Requirements brief\ModelTag}\\Isolated LLM call ($\le 8$ obligations)};
      \node[stage] (rplan) at (14.25,-1.05)
      {\textbf{Coverage plan\ModelTag}\\Isolated LLM call: prose research plan};

      \node[stage] (oscout) at (8.4,-2.08)
      {\textbf{Blind obligation scout\ModelTag}\\Request only; $\le 8$ additions to brief};
      \node[stage] (rscout) at (14.25,-2.08)
      {\textbf{Blind obligation scout\ModelTag}\\Request only; $\le 8$ additions to plan};
      \draw[flow] (obrief) -- (oscout);
      \draw[flow] (rplan) -- (rscout);

      \node[research,minimum height=.88cm] (bresearch) at (2.55,-3.18)
      {\textbf{Research and draft}\\Hybrid RRF ($\kappa=60$), $k=10$; adj.\ on\\Shared loop (100-round cap)};
      \node[research,minimum height=.88cm] (oresearch) at (8.4,-3.18)
      {\textbf{Research and draft}\\Hybrid RRF ($\kappa=60$), $k=10$; adj.\ on\\Shared loop (100-round cap)};
      \node[research,minimum height=.88cm] (rresearch) at (14.25,-3.18)
      {\textbf{Research and draft}\\Semantic + keyword, $k=10$; adj.\ on\\Separate research loop (40-round cap)};
      \draw[flow] (bbrief) -- node[note,fill=white,text width=4.5cm] {No blind scout} (bresearch);
      \draw[flow] (oscout) -- (oresearch);
      \draw[flow] (rscout) -- (rresearch);

      \node[stage,minimum height=.98cm] (breview) at (2.55,-4.45)
      {\textbf{Draft review (once)}\\Uncited scan\CodeTag; reviewer\ModelTag\\Checks draft against requirements brief};
      \node[stage,minimum height=.98cm] (oreview) at (8.4,-4.45)
      {\textbf{Draft review (once)}\\Uncited scan\CodeTag; reviewer\ModelTag\\Brief checked; scout items excluded};
      \draw[flow] (bresearch) -- (breview);
      \draw[flow] (oresearch) -- (oreview);
      \draw[conditional] (breview.east) -- (5.15,-4.45)
      -- node[note,rotate=90,below,inner sep=1pt] {feedback} (5.15,-3.18) -- (bresearch.east);
      \draw[conditional] (oreview.east) -- (11,-4.45)
      -- node[note,rotate=90,below,inner sep=1pt] {feedback} (11,-3.18) -- (oresearch.east);

      \node[stage,text width=16.15cm,minimum height=.48cm] (export) at (8.4,-5.52)
      {\textbf{Output export\CodeTag:} map \texttt{<docid>\_p<page>} citations to parent \texttt{docid} and emit RAG26 JSONL};
      \draw[flow] (breview.south) -- node[note,right] {no feedback} (breview.south |- export.north);
      \draw[flow] (oreview.south) -- node[note,right] {no feedback} (oreview.south |- export.north);
      \draw[flow] (rresearch.south) -- node[note,fill=white,text width=4.5cm]
      {No post-draft review\\Final answer from the same writer loop}
      (rresearch.south |- export.north);
      \draw[conditional] (bresearch.west) -- (-.15,-3.18)
      -- node[note,rotate=90,above,inner sep=1pt] {revised draft} (-.15,-5.08)
      -- (1.3,-5.08) -- (1.3,0 |- export.north);
      \draw[conditional] (oresearch.west) -- (5.7,-3.18)
      -- node[note,rotate=90,above,inner sep=1pt] {revised draft} (5.7,-5.08)
      -- (7.15,-5.08) -- (7.15,0 |- export.north);

      \node[draw=mainblockcolor!80!black,fill=mainblockcolor!7,line width=.7pt,rounded corners=2pt,
        minimum width=16.8cm,minimum height=2.55cm] at (8.4,-7.35) {};
      \node[heading] at (8.4,-6.33) {Expansion of each purple Research and draft block (dashed orange = P; solid blue = L/C)};
      \node[prompt,text width=5.65cm,minimum height=.88cm] (model) at (3.6,-7.18)
      {\textbf{Research/writer instructions\PromptTag}\\Formulate queries; assess coverage gaps;\\draft cited prose in the same conversation};
      \node[stage,text width=6.15cm,minimum height=.88cm] (tools) at (13.35,-7.18)
      {\textbf{Retrieval tools and evidence ledger\CodeTag}\\Search / fetch; expand $\pm 1$ adjacent chunks (top 5);\\stage batch; retain up to 10 via \texttt{commit\_context}};
      \draw[flow] (model.east) ++(0,.18) -- node[note,above] {tool calls} ([yshift=.18cm]tools.west);
      \draw[flow] (tools.west) ++(0,-.18) -- node[note,below] {staged / committed chunks} ([yshift=-.18cm]model.east);
      \node[note,text width=16cm] at (8.4,-8.12)
      {Unselected staged chunks expire next turn. Code enforces a 500k-token context budget, valid citation IDs, at most three citations/sentence, and a 1,024-word ceiling (no minimum). No code-level entailment checks.};
    \end{tikzpicture}%
  } \caption{Control flows for three submissions: our leading coverage run
  (\texttt{brief-hybrid}) and two SxS priority runs (\texttt{owa-glm5},
  \texttt{sol-research-first}). Purple blocks are expanded in the panel below.
  Solid arrows show execution order, not prompt inputs; each scout sees only the
  request. Unboxed notes mark absent stages; $k$ is the typical requested
  passages per search. Dashed blue arrows are conditional: review feedback
  resumes the writer loop, which may search again; the revised draft bypasses
  the once-only review. For staged evidence see
  Section~\ref{sec:trajectory-training}.} \Description{Three separate vertical
  pipelines for brief-hybrid, owa-glm5, and sol-research-first. Brief-hybrid
  uses GPT-5.6-Sol in all roles, with brief and review stages, hybrid RRF
  retrieval at k=10, and one review pass. Owa-glm5 uses GLM-5 across all stages
  and adds a blind obligation scout whose items are excluded from the review
  hook. Sol-research-first uses GPT-5.6-Sol in all roles with a prose coverage
  plan, blind scout, semantic plus keyword search at k=10, and a final answer
  from the same 40-round research loop. No blind scout for brief-hybrid and no
  post-draft review for sol-research-first are unboxed absence notes, not
  processing stages. Solid arrows show execution order; scouts receive only the
  request, not the preceding stage's output. Brief, scout, and reviewer stages
  are separate LLM calls; uncited scans and export are code. Conditional arrows
  return review feedback to the writer, which may search again, and send revised
  drafts directly to export without another review. Purple fill links the three
  research-and-draft blocks to a lower panel expanding their shared
  staged-evidence loop with adjacent-chunk expansion, commit_context retention,
  500k context limit, and code-enforced citation and word limits.}
  \label{fig:submitted-pipelines}
\end{figure*}

\subsection{End-to-End Architectures}

We use component markers to distinguish instructions embedded in
prompts~\PromptTag, separately invoked LLM stages~\ModelTag, and executable
code~\CodeTag.

All seven submissions generate cited reports of up to 1,024 words directly from
the test topics. Citations to individual chunks are mapped to parent document
IDs in the final output. Table~\ref{tab:run-configurations} summarizes the
system configurations, while Figure~\ref{fig:submitted-pipelines} illustrates
the control flow of our leading coverage run and two SxS priority runs.

\begin{table*}[bt]
      \centering
      \caption{Configurations of the submitted systems. Runs marked $\star$
            entered NIST human SxS assessment. Models denote the main and
            supporting LLMs (orchestrator and analyzer for \texttt{facetrag}).
            Sem: semantic; Kw: keyword; Hyb: hybrid RRF; SSR: shortest-substring
            ranking; Bool: Boolean search; Brief: requirements brief; Scout:
            blind obligation scout; Rev: post-draft review; Adj.: automatic
            adjacent-chunk expansion. Retrieval $k$ denotes the most frequently
            requested number of passages per search call.}
      \label{tab:run-configurations}
      \setlength{\tabcolsep}{4pt}
      \small
      \begin{tabular}{@{}lllp{4.1cm}c@{}}
    \toprule
    System                              & Models (main / support)       & Retrieval ($k$)                & Planning / Review         & Adj. \\
    \midrule
    \texttt{sol-default}                & Sol / ---                     & Sem+Kw (10)                    & None                      & No   \\
    \texttt{sol-research-first}$^\star$ & Sol / Sol                     & Sem+Kw (10)                    & Plan+Scout                & Yes  \\
    \texttt{brief-base}                 & Sol / Sol                     & Sem+Kw (10)                    & Brief+Rev                 & Yes  \\
    \texttt{brief-hybrid}               & Sol / Sol                     & Hyb (10)                       & Brief+Rev                 & Yes  \\
    \texttt{facets}                     & Luna / ---                    & Sem+Kw (10), Hyb (15)          & Facets+coverage check     & No   \\
    \texttt{facetrag}                   & GPT-OSS-120B / Qwen3-Next-80B & Sem+Kw+SSR+Bool (10), Hyb (15) & Facets+multi-stage review & ---  \\
    \texttt{owa-glm5}$^\star$           & GLM-5 / GLM-5                 & Hyb (10)                       & Brief+Scout+Rev           & Yes  \\
    \bottomrule
\end{tabular}

\end{table*}

Retrieval $k$ is selected by the agent for each search call, except in
\texttt{facetrag}, where it is fixed by code. When adjacent-chunk expansion is
enabled, the system automatically retrieves the preceding and following chunks
for the top-ranked search results.

Five submissions share a staged-evidence research architecture, differing in
models, retrieval methods, planning stages, and review mechanisms. A shared
evidence ledger lets agents select and retain retrieved chunks for report
generation. Code enforces a 500,000-token context budget, a maximum of three
citations per sentence, and the report word limit.

\begin{itemize}
      \item \textbf{\texttt{sol-default}} uses a single \texttt{GPT-5.6-Sol}
            conversation for research and drafting, with up to 100 research
            rounds\PromptTag and no separate planning or review stages.

      \item \textbf{\texttt{sol-research-first}} uses \texttt{GPT-5.6-Sol}
            across all stages. An isolated coverage planner\ModelTag generates a
            research plan, supplemented by up to eight additional requirements
            from a blind obligation scout\ModelTag that sees only the request.
            These outputs then guide a 40-round research loop, followed by
            drafting and a single-pass post-draft review\PromptTag.

      \item \textbf{\texttt{brief-base}} and \textbf{\texttt{brief-hybrid}}
            extend \texttt{sol-default} with an isolated pre-flight requirements
            brief\ModelTag and a single-pass post-draft review\ModelTag, both
            using \texttt{GPT-5.6-Sol}. The writer conducts research and
            drafting, guided by the brief\PromptTag. A code-based citation
            scan\CodeTag flags uncited sentences, and reviewer feedback\ModelTag
            is returned to the writer for revision. The two runs differ
            primarily in retrieval: \texttt{brief-base} provides the agent with
            semantic and keyword search, while \texttt{brief-hybrid} uses hybrid
            retrieval with Reciprocal Rank Fusion (RRF).

      \item \textbf{\texttt{owa-glm5}} follows the \texttt{brief-hybrid}
            architecture but uses the open-weight \texttt{GLM-5} model across
            all stages. It also adds a blind obligation scout\ModelTag, whose
            requirements supplement the initial brief but are excluded from the
            post-draft review.
\end{itemize}

Pre-flight briefs and blind scouts each identify up to eight answer
requirements. Scouts receive only the original test topic, independently of the
other planning stages. Two additional submissions explore explicit facet
decomposition, dividing each topic into subquestions requiring separate
evidence:

\begin{itemize}
      \item \textbf{\texttt{facets}} uses a single \texttt{GPT-5.6-Luna}
            conversation to plan, research, and draft across multiple
            facets\PromptTag. Each search is associated with a specific
            requirement. A code-based coverage check\CodeTag identifies
            unresolved requirements and may return the draft for further
            research. No separate reviewer is used.

      \item \textbf{\texttt{facetrag}} combines a \texttt{GPT-OSS-120B}
            orchestrator with a \texttt{Qwen3-Next-80B-A3B} analyzer to research
            3--6 facets concurrently. Each facet has separate search-planning
            calls\ModelTag and supports semantic, keyword, hybrid, and Boolean
            retrieval, including Shortest-Substring Ranking (SSR). The
            analyzer\ModelTag identifies relevant passages and remaining gaps.
            An LLM curator\ModelTag selects evidence based on relevance and
            diversity using an MMR-style approach~\cite{carbonell1998mmr},
            supplying up to five items per facet. Separate drafting,
            citation-checking, and formatting stages\ModelTag produce the final
            report, followed by code validation\CodeTag.
\end{itemize}

\subsection{Factorization of Design Choices}

RAG systems combine interacting choices across retrieval, generation, curation,
and evaluation; motivating systematic comparison of individual pipeline
components~\cite{rau2024bergen}. We catalogued 14 structural factors and
evaluated selected ablations (Section~\ref{sec:results}), alongside writer and
helper model choices (writer, brief analyst, reviewer).\footnote{Detailed factor
definitions, implementation pointers, and reports are available in the code
repository: \url{https://github.com/rmit-ir/trec-rag-26}.} Factors include
pre-flight briefing, blind obligation scouting, single-pass review, $\pm 1$
adjacent-chunk expansion, retrieval engines and cutoffs, search-preview
truncation, full-chunk staging versus LLM-generated snippets, and evidence
pruning. \texttt{brief-base} and \texttt{brief-hybrid} were developed using
one-factor-at-a-time hill-climbing on a fixed 15-topic development set,
evaluated with a rubric grader and head-to-head arena comparisons.
\texttt{brief-hybrid} differs only in its retrieval engine (hybrid only). It
achieved the highest rubric score, confirmed on a second 15-topic set, but a
lower arena win rate than \texttt{brief-base}, so both were submitted.

\texttt{sol-research-first} was independently developed on a 30-topic
development set against a Sol baseline. Adding an isolated coverage plan and a
blind atomic scout to a continuous research conversation, which searches,
retains evidence, and drafts the answer, improved rubric scores. Adding a
post-draft repair stage reduced performance, so this stage was dropped.
``First'' refers to the ordering of planning and research before drafting.
Across all runs, code enforces context, round, length, and citation-ID
constraints, while semantic completeness and citation entailment remain
model-driven.

\subsection{Resource Usage and Cost Estimates}
\label{sec:runs}

\begin{table}[t]
      \centering

      \caption{Average resource usage and estimated cost per report across all
            119 test topics. Srch: search calls; Refs: cited references; In:
            uncached input tokens; Cache: cached input tokens; Out: output
            tokens. Token counts are in thousands (K). Runs marked $\star$
            entered NIST human SxS assessment.}

      \label{tab:submissions}
      \resizebox{\columnwidth}{!}{
            \begin{tabular}{@{}lrrrrrrr@{}}
    \toprule
    System                              & Srch & Words & Refs & In (K) & Cache (K) & Out (K) & \$/topic \\
    \midrule
    \texttt{sol-default}                & 23.9 & 882.5 & 27.9 & 182.2  & 202.4     & 6.1     & 1.566    \\
    \texttt{sol-research-first}$^\star$ & 31.4 & 936.6 & 27.7 & 330.5  & 369.3     & 10.5    & 2.823    \\
    \texttt{brief-base}                 & 26.4 & 830.2 & 25.8 & 319.6  & 388.5     & 8.0     & 2.356    \\
    \texttt{brief-hybrid}               & 22.6 & 834.3 & 24.7 & 293.1  & 322.4     & 7.5     & 2.156    \\
    \texttt{facets}                     & 11.7 & 613.0 & 13.6 & 138.9  & 71.4      & 5.9     & 0.040    \\
    \texttt{facetrag}                   & 33.7 & 388.9 & 13.9 & 411.7  & 0         & 49.7    & 0.106    \\
    \texttt{owa-glm5}$^\star$           & 12.7 & 754.0 & 15.9 & 415.1  & 0         & 5.9     & 0.524    \\
    \bottomrule
\end{tabular}

      }
\end{table}

Table~\ref{tab:submissions} summarizes resource usage and estimated costs across
all 119 test topics. \texttt{sol-research-first} produced the longest reports
(936.6 words on average) and consumed 81\% more processed tokens than
\texttt{sol-default}. Compared with \texttt{brief-base}, \texttt{brief-hybrid}
maintained similar report length and citation counts while reducing searches by
14.4\% and processed tokens by 8.2\%. The open-weight \texttt{owa-glm5} required
fewer searches (12.7 per topic) and had a substantially lower estimated cost
than \texttt{sol-research-first} (\$0.524 vs.\ \$2.823 per topic).
\texttt{facetrag} issued the most searches (33.7 per topic) but produced the
shortest reports (388.9 words on average), despite generating 49.7K output
tokens across its multi-stage pipeline.

Costs are averages over successful attempts on the 119 topics, calculated from
traced input, cache-read, and output tokens, supplemented by logged brief,
review, and scout calls where necessary, using provider-specific rates.
Estimates cover generation only, excluding failed attempts, experimentation, and
judging.

\section{Retrieval and Evidence Handling}
\label{sec:retrieval}

We segmented \texttt{ClimbMix-400B-shuffle}~\cite{diao2025climb} (6,543
shards, approximately 563 million documents) into paragraph-aware chunks, following recent findings that simple structure-based chunking strategies perform well for in-corpus retrieval~\cite{zhou2026beyond}. The
resulting BM25 index contains 921,892,634 chunks. Token counts were estimated at
1.3 tokens per word, targeting 200--500 tokens per chunk with a hard cap of 700.
Each chunk was assigned a unique, 1-based \texttt{<docid>\_p<page>} identifier.

\textbf{Sparse Lexical and Boolean Retrieval.}
We used a self-hosted Lucene/Anserini
index~\cite{yang2017anserini,lin2021pyserini} with Pyserini's default BM25
parameters ($k_1=0.9$, $b=0.4$) for keyword and Lucene Boolean search.
\texttt{facetrag} additionally supports Shortest-Substring Ranking
(SSR)~\cite{clarke2000ssr} through a Cottontail-based Boolean retrieval
system~\cite{clarke2026boolean}, which ranks results by compact matching spans.

\textbf{Dense Semantic Retrieval.}
For semantic retrieval, we used our
\texttt{jina-v5-nano-trecrag26-agent-256d}
model,\footnote{Model card:
    \url{https://huggingface.co/RMIT-ADMS/jina-v5-nano-trecrag26-agent-256d}.}
    fine-tuned from
    \texttt{jina-embeddings-v5-text-nano}~\cite{akram2026jinav5}, producing
    256-dimensional embeddings. We indexed these embeddings using
    DiskANN~\cite{subramanya2019diskann}, an approximate nearest-neighbor search
    method designed for large-scale, disk-based retrieval. The index used
    maximum inner product search, with graph degree $R=64$, construction search
    list size $L_{\text{build}}=100$, and 32-byte product quantization.

On 22 TREC RAG 2025 development topics (condensed metrics averaged over three
judge variants), the fine-tuned model achieved slightly higher aggregate scores
than the base model: nDCG@10 (0.720 vs.\ 0.713), MAP (0.113 vs.\ 0.103), and
R@1000 (0.116 vs.\ 0.106). These small differences should be interpreted
cautiously given the limited validation set.

\textbf{Hybrid Retrieval.}
Hybrid search combines semantic and lexical retrieval using unweighted
Reciprocal Rank Fusion (RRF)~\cite{cormack2009rrf}, with $\kappa=60$. Both
retrievers return at least 50 candidates before fusion, and the top-$k$ chunks
are returned. When adjacent-chunk expansion is enabled
(Table~\ref{tab:run-configurations}), the system also retrieves the preceding
and following chunks for the top five results, removing duplicates.

\textbf{Query Generation.}
For hybrid retrieval, \texttt{facets} and \texttt{facetrag} use HyDE-style
instructions~\cite{gao2023hyde} that encourage generating short hypothetical
answer passages as search queries. Query formulations can vary substantially
across both generation strategies and search contexts
~\cite{zendel2025llmquerydiversity,abu2026heterogeneous}, motivating direct
evaluation of the query-generation strategy. These instructions appear in the
main prompt for \texttt{facets} and in individual search-planning calls for
\texttt{facetrag}. They do not enforce this query format, and hypothetical
passages are used only for retrieval, not as evidence.

On 15 development topics, plain hybrid retrieval scored 2.333/3, compared with
2.267/3 when adding HyDE-style instructions. Given the small difference, we
found no clear benefit from HyDE-style query generation and used plain hybrid
retrieval in \texttt{brief-hybrid} and its \texttt{owa-glm5} derivative.

\subsection{Staged Evidence and Retriever Fine-Tuning}
\label{sec:trajectory-training}

We used a staged evidence approach to manage retrieved chunks and collect
training data for the fine-tuning of the retriever. Retrieved chunks are
initially available for one turn. The agent can call \texttt{commit\_context} to
retain selected chunks in its evidence ledger; uncommitted chunks are reduced to
markers and expire from the staged batch. We fine-tuned our dense retrieval
encoder using evidence-selection decisions from initial agent runs. Recent work
has similarly shown that search-agent trajectories can provide useful
supervision for training retrievers that account for the agent's interaction
history~\cite{zhou2026lrat,chen2026iter}. Two \texttt{GPT-5.6- \allowbreak Luna}
configurations (semantic-only and keyword-only) processed all 119 official test
topics. Chunks committed at any point during a run were labeled positive, while
retrieved but non-selected chunks were treated as negative candidates. The
resulting dataset export contained 2,143 search rows (1,073 semantic and 1,070
keyword), with 4,857 positive and 12,889 negative query--chunk pairs. These
counts describe the exported data, not the final training set.

We trained the retriever on 95 topics, comprising 1,587 search sub-queries and
3,849 positive query--chunk pairs, and held out 24 topics for validation. We
used a patched version of Tevatron~\cite{gao2023tevatron} with
InfoNCE~\cite{oord2018cpc}, contrasting positive query--chunk pairs against
other passages in the same batch (in-batch negatives). Training used
$\tau=0.05$, batch size 64, 4 epochs, learning rate $10^{-4}$, bf16, and the
first 256 embedding dimensions before normalization. We applied LoRA with
$r=16$, $\alpha=64$, and dropout $0.1$. On the 24 held-out topics (410
sub-queries), we observed an improvement in nDCG@10 from 0.751 to 0.793,
evaluated over a pool of 4,071 agent-committed chunks. Based on agent-derived
labels, rather than human relevance judgments or full-corpus retrieval.

The training data came from the official 2026 test topics, without using actual
relevance labels. The held-out retrieval evaluation therefore does not establish
independent end-to-end generation performance.\footnote{Training details and
reproducibility limitations are documented at
\url{https://github.com/rmit-ir/trec-rag-26/blob/main/docs/embedding-training.md}.}
All dense and hybrid submissions used this fine-tuned encoder.

\section{Internal Evaluation (Pre-Submission)}
\label{sec:results}

We report pre-submission evaluations separately from the official results
(Section~\ref{sec:official}). These evaluations used different assessment
methods and, in some cases, different topic sets, so their scores are not
directly comparable with the organizers' results. Nevertheless, they provide
insights into the relative performance of our systems under automated evaluation
and the effects of design choices, helping guide our final submissions.

\subsection{Evaluation Protocols and Limitations}
\label{sec:setup}

Pre-submission evaluation used two automated RAGDoll protocols: pointwise
grading and pairwise arena comparisons.\footnote{See
\url{https://github.com/castorini/RAGDoll}.} Pointwise grading assessed
approximately 31 criteria (scored 0--2) across six dimensions (Communication,
Explicit Criteria, Implicit Criteria, Instruction Following,
References/Citations, and Synthesis), plus an overall quality score (0--3),
using \texttt{GPT-5.6-Terra}. Pairwise comparisons assessed which of two
responses was preferred, evaluating both presentation orders (AB/BA).

System development used the 30 organizer
Research-Rubrics~\cite{sharma2025researchrubrics} development topics, split into
two disjoint 15-topic sets for tuning and validation. No submitted system
configuration was selected using the 119 test topics, although the retriever was
separately fine-tuned on agent trajectories from test topics
(Section~\ref{sec:trajectory-training}).

Our internal experiments revealed substantial sensitivity to the evaluation
setup. On the same 20 topics, changing only the pairwise judge shifted our
system's win rate against the organizer baseline from 17.5\%
(\texttt{GPT-5.6-Sol}) to 45.0\% (\texttt{GPT-5.6-Luna}), with
\texttt{GPT-5.6-Terra} giving 25.0\%. The judges were not equally independent of
the compared systems, and changing answer presentation order also affected
outcomes. Arena preferences were correlated with response length, while separate
citation-support checks identified weaknesses that pairwise judgments did not
capture. These findings highlight the limitations of relying on any automated
judgments alone, consistent with prior observations that LLM-based assessment
can be sensitive to evaluator and presentation
choices~\cite{dietz2025llmjudges,zendel2024annotation}.

\subsection{Writer, Retrieval, and Orchestration Sweeps}
\label{sec:ablations}

In the 15-topic \texttt{brief-base} sweep, writer choice had the largest impact
on pointwise quality. Relative to the \texttt{GPT-5.6-Sol} baseline (2.267/3),
switching to \texttt{GPT-5.6-Terra} or \texttt{Luna} reduced scores by 0.40,
\texttt{Qwen3-Next-80B-A3B} by 0.867, and \texttt{GPT-OSS-120B} by 1.067. Of 12
structural changes, 10 changed scores by at most 0.067 (equivalent to one topic
on this benchmark), while two reduced scores by 0.133 (search-preview truncation
and adding SSR with Lucene Boolean search). Hybrid-only RRF
(\texttt{brief-hybrid}) improved the pointwise score (2.333 vs.\ 2.267), with a
larger advantage on the second, 15-topic validation set (2.333 vs.\ 2.133). As
shown in Table~\ref{tab:submissions}, across the 119 test runs,
\texttt{brief-hybrid} used 14.4\% fewer searches and 8.2\% fewer processed
tokens than \texttt{brief-base}, with 8.5\% lower estimated generation cost
(\$2.156 vs.\ \$2.356 per topic). However, its 15-topic arena win rate against a
Luna-based \texttt{sol-research-first} decreased from 40.0\% to 33.3\%, using a
Terra judge across both presentation orders (30 battles).

Disabling $\pm1$ adjacent-chunk expansion reduced pointwise scores by 0.067 with
Sol and 0.133 with Qwen3. Among development-time runs of the corresponding
architectures on the 15-topic benchmark, \texttt{brief-base} (Sol) and
\texttt{sol-research-first} (Luna) tied at 2.267/3, ahead of
\texttt{sol-default} (Luna, 2.000) and \texttt{facets} (1.800). In a pairwise
arena, \texttt{brief-base} lost to \texttt{sol-research-first} by 12 wins to 18
across 30 battles (15 topics evaluated in both orders). On a separate 30-topic
development rubric, the \texttt{sol-research-first} configuration scored higher
than \texttt{sol-default} (0.704 vs.\ 0.651). Across the report's 30
proprietary-model cells, \emph{References and Citation Quality} was the weakest
dimension (mean 0.167/2, range 0.00--0.67). The \texttt{sol-research-first}
development run scored 0.0 on this dimension, although this was based on only
nine rubric criterion instances.

Overall, our internal experiments suggest that writer choice had a greater
impact than most structural changes, while hybrid retrieval offered modest
quality and efficiency improvements. The evidence-handling analysis also
highlight a trade-off between token efficiency and citation support, and show
that automated preference judgments can be sensitive to judge choice and may
overlook citation-support weaknesses. However, these findings are based on a
limited number of topics and should be interpreted cautiously. They do not
necessarily predict performance on the official evaluation, which used different
topics and assessment methods. Further research is needed to understand the
generalizability of these results.

\subsection{Open-Weight Analysis and Baseline Arena}
\label{sec:open-weight-diagnostics}

On two 15-topic development sets, \texttt{owa-glm5} scored 1.467/3 and 1.333/3,
below the proprietary \texttt{GPT-5.6} configurations (1.867--2.333). The
largest gaps were in Implicit Criteria ($-0.294$) and Synthesis ($-0.205$).
However, \texttt{owa-glm5} scored higher on References and Citations (0.667/2
vs.\ 0.222 and 0.0), suggesting that its more frequent citation use may benefit
this rubric dimension.

In an attempt to isolate the effects of the writer from retrieval and
orchestration, we replaced the \texttt{GPT-5.6-Sol} writer in the
\texttt{sol-research-first} system with \texttt{Qwen3-Next-80B-A3B}, keeping the
same retriever and evidence collection strategy. Replaying evidence collected by
\texttt{sol-research-first} into Qwen3 improved Synthesis ($+0.154$) but barely
changed Implicit Criteria ($+0.006$), yielding an overall score of 1.267.
Replacing Qwen3 helper models with \texttt{GPT-5.6-Luna} slightly reduced
performance (1.0--1.067 vs.\ 1.20). Our internal experiments suggest that writer
synthesis, rather than evidence collection or helper model quality alone, seem
to be an important bottleneck.

To estimate performance in the official evaluation, we compared six completed
runs against the organizer's Sol agentic-search BM25 baseline on all 119 test
topics using automated preference judgments. Unlike the development topics,
these had no available evaluation labels or rubric, , so we used the arena as a
pre-submission sanity check rather than for system selection or optimization.
\texttt{owa-glm5} was finalized after this sweep and was not included.
Table~\ref{tab:organizer-arena} summarizes the results.

\begin{table}[tb]
    \centering
    \caption{Internal both-order RAGDoll arena comparisons against the
        organizer's shipped agentic BM25 baseline over all 119 narratives
        (GPT-5.6-Luna judge, seed 13; not official TREC judgments). Win rate is
        $(W + 0.5T)/119$. \texttt{owa-glm5} was finalized after this sweep.}
    \label{tab:organizer-arena}
    \setlength{\tabcolsep}{5pt}
    \begin{tabular}{@{}lrrrr@{}}
    \toprule
    System                      & W   & L   & T   & Win rate \\
    \midrule
    \texttt{sol-research-first} & 108 & 10  & 1   & 91.2\%   \\
    \texttt{sol-default}        & 107 & 11  & 1   & 90.3\%   \\
    \texttt{brief-hybrid}       & 101 & 18  & 0   & 84.9\%   \\
    \texttt{brief-base}         & 99  & 19  & 1   & 83.6\%   \\
    \texttt{facets}             & 13  & 105 & 1   & 11.3\%   \\
    \texttt{facetrag}           & 0   & 119 & 0   & 0.0\%    \\
    \texttt{owa-glm5}           & --- & --- & --- & ---      \\
    \bottomrule
\end{tabular}

\end{table}

In the arena against the organizer's baseline results all four Sol runs achieved
high win rates (83.6--91.2\%), while \texttt{facets} (11.3\%) and
\texttt{facetrag} (0.0\%) performed poorly. Internal pairwise comparisons
revealed observable position bias. \texttt{sol-research-first} narrowly led
\texttt{sol-default} (67 wins, 51 losses, 1 tie), but its non-tie win rate
changed from 37.7\% when presented as A to 76.7\% as B. Similarly,
\texttt{brief-hybrid} and \texttt{brief-base} were nearly tied overall (58 wins,
57 losses, 4 ties), despite win rates changing from 34.5\% to 63.9\% between
display orders.

Compared with official NIST coverage results in
Table~\ref{tab:official-coverage}, the internal arena broadly distinguished
stronger and weaker systems but did not reliably reproduce their relative
rankings. Across the six common systems, the arena's ranking agreed only
moderately with the official coverage rankings (Kendall's $\tau=0.33$-$0.60$;
3-5 of 15 system pairs ordered differently; none significant at $n=6$), with the
highest agreement for B1 all-items and B2 automatic nuggets ($\tau=0.60$). For
example, \texttt{sol-default} nearly matched \texttt{sol-research-first}
internally despite substantially lower B1 vital coverage (0.790 vs.\ 0.892).
Together with the observed judge and position biases, these discrepancies
highlight limitations of automated evaluation and motivate initiatives such as
TREC Auto-Judge~\cite{farzi2026autojudge}, which investigate the reliability of
automated assessments.

\section{Official TREC Evaluation}
\label{sec:official}

\begin{table*}[t]
    \centering
    \caption{Official NIST TREC 2026 RAG coverage (\texttt{vital\_strict} /
        \texttt{all\_strict}) and numbers of narratives scoring at or above the
        global track median ($\ge$Med, V/A) across our seven submitted runs.
        Underlining marks our highest run per metric; $\star$ denotes a tie with
        the global track maximum before rounding. Two B2 manual narratives
        (\texttt{rag2026-89}, \texttt{rag2026-114}) contain no vital nuggets,
        capping \texttt{vital\_strict} at $23/25=0.920$.}
    \label{tab:official-coverage}
    \small
    \setlength{\tabcolsep}{5pt}
    \begin{tabular}{@{}llcccccc@{}}
    \toprule
                                 &                             & \multicolumn{2}{c}{B1 Human Rubric ($n=17$)}  & \multicolumn{2}{c}{B2 Manual Nuggets ($n=25$)} & \multicolumn{2}{c}{B2 Auto Nuggets ($n=25$)}                                                               \\
    \cmidrule(lr){3-4} \cmidrule(lr){5-6} \cmidrule(lr){7-8}
    Type                         & System                      & Vital / All                                   & $\ge$Med (V/A)                                 & Vital / All                                  & $\ge$Med (V/A) & Vital / All               & $\ge$Med (V/A) \\
    \midrule
    \multirow{5}{*}{Proprietary} & \texttt{brief-hybrid}       & \underline{0.892}$^\star$ / \underline{0.909} & 17 / 17                                        & \underline{0.691} / \underline{0.723}        & 24 / 25        & 0.644 / \underline{0.565} & 25 / 25        \\
                                 & \texttt{sol-research-first} & \underline{0.892}$^\star$ / 0.887             & 17 / 17                                        & 0.684 / 0.702                                & 25 / 24        & \underline{0.649} / 0.560 & 25 / 25        \\
                                 & \texttt{facets}             & 0.831 / 0.841                                 & 17 / 16                                        & 0.627 / 0.642                                & 23 / 23        & 0.390 / 0.333             & 25 / 24        \\
                                 & \texttt{brief-base}         & 0.824 / 0.862                                 & 17 / 16                                        & 0.691 / 0.721                                & 24 / 25        & 0.648 / 0.555             & 25 / 25        \\
                                 & \texttt{sol-default}        & 0.790 / 0.861                                 & 17 / 17                                        & 0.679 / 0.705                                & 25 / 25        & 0.630 / 0.563             & 25 / 25        \\
    \midrule
    \multirow{2}{*}{Open-weight} & \texttt{owa-glm5}           & 0.853 / 0.821                                 & 17 / 16                                        & 0.550 / 0.605                                & 23 / 23        & 0.288 / 0.243             & 22 / 23        \\
                                 & \texttt{facetrag}           & 0.600 / 0.622                                 & 11 / 10                                        & 0.460 / 0.481                                & 13 / 10        & 0.139 / 0.110             & 12 / 11        \\
    \midrule
                                 & \emph{TREC Track Max}       & \emph{0.892 / 0.911}                          & ---                                            & \emph{0.729 / 0.780}                         & ---            & \emph{0.666 / 0.584}      & ---            \\
                                 & \emph{TREC Track Median}    & \emph{0.716 / 0.733}                          & ---                                            & \emph{0.488 / 0.556}                         & ---            & \emph{0.155 / 0.128}      & ---            \\
                                 & \emph{TREC Track Min}       & \emph{0.039 / 0.043}                          & ---                                            & \emph{0.076 / 0.070}                         & ---            & \emph{0.002 / 0.002}      & ---            \\
    \bottomrule
\end{tabular}

\end{table*}

\subsection{Rubric and Nugget Coverage}
\label{sec:coverage-analysis}

Table~\ref{tab:official-coverage} reports official \texttt{vital\_strict} and
\texttt{all\_strict} coverage~\cite{trec2026rag,pradeep2024autonuggetizer,
pradeep2025nuggetrecall,voorhees2003overview} across Batch-1 human-intuition
rubrics ($n=17$), Batch-2 manually authored nuggets ($n=25$), and Batch-2
automatically generated RAGDoll AutoNuggetizer nuggets ($n=25$). In all three
batches, item support was assigned automatically using \texttt{GPT-6.1 Sol}
at medium reasoning. Strict scoring gives credit only when an item is fully
supported; partial support receives zero credit. \texttt{vital\_strict}
considers vital items only, whereas \texttt{all\_strict} includes vital and
``okay'' items with equal weight.

Our strongest coverage results came from \texttt{brief-hybrid} and
\texttt{sol-research-first}. Both tied the global track maximum on Batch-1
\texttt{vital\_strict} at 0.892, while \texttt{brief-hybrid} achieved our
highest \texttt{all\_strict} score in all three batches (0.909, 0.723, and
0.565). Six of seven submitted runs exceeded the global track median across all
three batches.

The nominal rankings should be interpreted with a grain of salt. We conducted
pairwise comparisons using Tukey's HSD test~\cite{tukey1949comparing} on the
per-topic coverage scores and found no significant differences among the four
leading proprietary runs, \texttt{brief-hybrid}, \texttt{brief-base},
\texttt{sol-default}, and \texttt{sol-research-first} on any of the six coverage
metrics. \texttt{facetrag} was consistently the weakest system, while the
separation between the leading runs, \texttt{facets}, and \texttt{owa-glm5} was
largest under Batch-2 automatic nuggets. The official coverage results therefore
distinguish the strongest and weakest systems more clearly than they distinguish
among the leading proprietary configurations.

We also observed a positive association between answer length and coverage
within topics, increasing from a mean Spearman correlation of 0.42 in
Batch~1 to 0.72 under Batch-2 automatic nuggets. This relationship is
descriptive: answer length is confounded with system design and does not imply
that longer answers alone cause higher coverage, but it does suggest that the
fully automated Batch-2 evaluation may favor longer responses and does not
penalize verbosity, unlike typical human evaluation.

\subsection{Human Side-by-Side Preference}
\label{sec:human-preference}

\begin{figure}[t]
    \centering
    \includegraphics[width=\columnwidth]{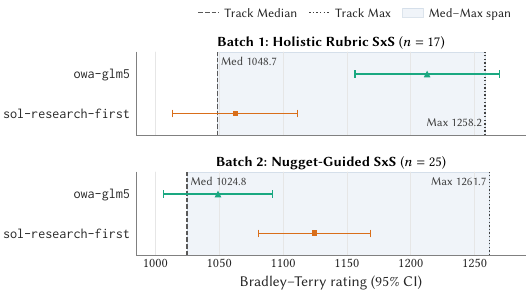}
    \caption{Official human side-by-side Bradley--Terry ratings with published
        95\% confidence intervals and track medians and maxima. Batch~1 ($n=17$)
        and Batch~2 ($n=25$) were fitted separately and used different topic
        sets and assessment protocols.} \Description{Two stacked panels showing
        Bradley--Terry ratings for owa-glm5 and sol-research-first. Owa-glm5
        leads in Batch 1, while sol-research-first leads in Batch 2.}
    \label{fig:official-preference}
\end{figure}

The organizers also performed human side-by-side evaluation for our two priority
runs, \texttt{owa-glm5} and \texttt{sol-research-first}, fitting separate
Bradley--Terry (BT) ratings for each batch~\cite{bradley1952rank}
(Figure~\ref{fig:official-preference}).

In Batch~1, assessors strongly preferred \texttt{owa-glm5}, with a BT rating of
1212.8 compared with 1062.7 for \texttt{sol-research-first}. The difference was
also robust when topics were treated as the unit of analysis: 13 of 17
topics favored \texttt{owa-glm5}, using a paired sign-flip test on
per-topic battle-score differences found the difference to be significant
with $p=0.006$. This contrasts with the coverage results, where
\texttt{sol-research-first} scored higher on both Batch-1 strict metrics,
showing that rubric coverage and holistic human preference can rank the same
systems differently. The direction reversed in Batch~2, where
\texttt{sol-research-first} achieved a higher BT rating (1124.7 vs.\ 1049.0).
However, the topic-level comparison was inconclusive ($p=0.110$), making
this result less robust than the Batch-1 preference. The batches also differed
in both topics and assessment protocol: Batch-1 assessors compared responses
holistically, whereas Batch-2 assessors first constructed document-grounded
nuggets before performing their comparisons.

Overall, the human judgments provide a complementary signal to coverage. In
particular, \texttt{owa-glm5}'s strong Batch-1 preference despite lower coverage
suggests that characteristics not captured by strict rubric recall influenced
assessor preferences. Section~\ref{sec:discussion} examines readability and
response style as possible contributors.

\begin{figure*}[t]
    \centering
    \includegraphics[width=\textwidth]{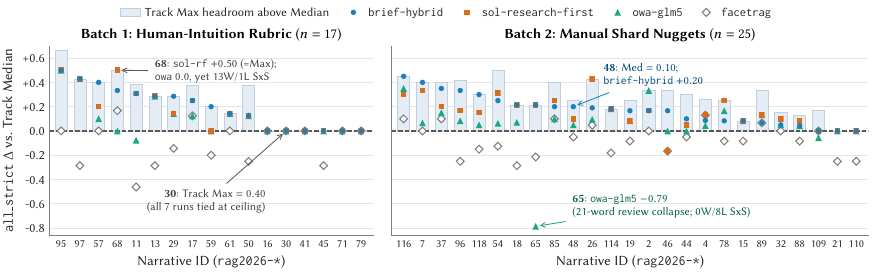}
    \caption{Per-topic official \texttt{all\_strict} coverage difference
        ($\Delta$) relative to the TREC 2026 RAG track median across Batch~1
        ($n=17$, left) and Batch~2 manual nuggets ($n=25$, right), sorted within
        each panel by \texttt{brief-hybrid} $\Delta$. Shaded bars show the
        difference between the track maximum and median.} \Description{Two-panel
        per-topic plot showing substantial variation in coverage relative to
        the track median, including the coverage-- preference divergence on
        rag2026-68 and the owa-glm5 failure on rag2026-65.}
    \label{fig:official-per-topic}
\end{figure*}

\subsection{Per-Topic Divergence}
\label{sec:failure-analysis}

Figure~\ref{fig:official-per-topic} shows substantial per-topic variation
and illustrates why aggregate coverage alone does not fully characterize system
behavior. We highlight two cases.

\textbf{Review-Pass Failure.}
On Batch-2 topic 65 (hiring-AI compliance), \texttt{brief-hybrid} and
\texttt{sol-research-first} both achieved 1.000/1.000 coverage, whereas
\texttt{owa-glm5} scored 0.0/0.0 and lost all eight side-by-side comparisons.
Its trace shows 12 searches and five evidence commits followed by a 922-word
draft. During review, the model was asked to add a citation to the opening
sentence while preserving the report, but returned only a corrected 21-word
sentence, which the harness accepted as the complete revision. This was
therefore a post-draft revision failure rather than a failure to retrieve or
synthesize evidence initially.

\textbf{Coverage--Preference Divergence.}
On Batch-1 topic 68 (school safety), \texttt{sol-research-first} achieved
0.750/0.833 coverage, tying the global maximum on \texttt{all\_strict}, but
recorded 6W/2T/6L in human comparisons. \texttt{owa-glm5}, despite substantially
lower coverage (0.250/0.333), recorded 13W/0T/1L. This example emphasizes the
surprising Batch-1 result: higher strict rubric coverage did not necessarily
imply stronger holistic human preference.

\section{Discussion}
\label{sec:discussion}

Batch~1 reveals a clear divergence between information coverage and holistic
human preference. Although \texttt{owa-glm5} covered fewer rubric items than
\texttt{sol-research-first} (Table~\ref{tab:official-coverage}), assessors
strongly preferred it in the side-by-side evaluation
(Figure~\ref{fig:official-preference}). This preference was robust at the
topic level, with 13 of 17 topics favoring \texttt{owa-glm5} (sign-flip
$p=0.006$), despite its lower Batch-1 coverage. The result is also notable from
an efficiency perspective: estimated generation cost was \$0.524 per topic for
\texttt{owa-glm5}, compared with \$2.823 for \texttt{sol-research-first}. The
per-topic example in Section~\ref{sec:failure-analysis} further illustrates
that higher strict coverage does not necessarily imply stronger human
preference.

\begin{table}[t]
\centering
\caption{Report characteristics on the 17 matched Batch-1 topics
(means, except topic counts).}
\label{tab:batch1-readability}
\small
\setlength{\tabcolsep}{4pt}
\begin{tabular}{@{}lrr@{}}
\toprule
Measure & \texttt{owa-glm5} & \texttt{sol-research-first} \\
\midrule
Words & 805.3 & 951.2 \\
Unique references & 16.6 & 28.9 \\
Flesch--Kincaid Grade Level $\downarrow$ & 15.7 & 22.2 \\
Topics with shorter report & 15/17 & 2/17 \\
\bottomrule
\end{tabular}
\end{table}

To examine whether reading burden might contribute to this difference, we
measured Flesch--Kincaid Grade Level~\cite{kincaid1975readability} on the 17
matched Batch-1 responses using \texttt{textstat} (US English).\footnote{Version
0.7.12: \url{https://github.com/textstat/textstat}.} We excluded citation IDs,
reference lists, and Markdown markers and used the library's sentence
segmentation. \texttt{owa-glm5} was easier to read on all 17 topics, with a mean
Flesch--Kincaid grade of 15.7 compared with 22.2 for \texttt{sol-research-first}
(sign test $p<0.001$; Table~\ref{tab:batch1-readability}). It also used shorter
sentences on average (21.8 vs.\ 34.9 words per sentence). However, the per-topic
readability difference was not associated with the per-topic preference margin
(Spearman $\rho=+0.27$, $p=0.30$, $n=17$), and no readability measure was
significant after correction for multiple comparisons. Answer length was
likewise not associated with the Batch-1 preference difference ($\rho=-0.15$,
$p=0.56$). Because \texttt{owa-glm5} was easier to read on every topic, these
data cannot separate readability from other systematic differences between the
systems, such as response structure, tone, or handling of uncertainty.
Readability therefore remains a plausible contributor rather than an explanation
of the observed preference.

The preference direction reversed in Batch~2, although the topic-level
difference was not statistically significant. Both the topics and assessment
protocol changed between batches, so this reversal cannot be attributed
specifically to nugget-guided assessment. More broadly, the divergence between
coverage and human preference, together with the judge sensitivity observed in
our internal evaluation (Section~\ref{sec:setup}), highlights the importance of
using complementary evaluation signals rather than relying on a single automated
or human metric.

\section{Conclusion}
\label{sec:conclusion}

\textbf{RQ1.} Open-weight agents were competitive on some measures but did not
consistently match the coverage of the leading proprietary systems.
\texttt{owa-glm5} exceeded the global track median across all official coverage
evaluations and was strongly preferred over \texttt{sol-research-first} in
Batch-1 human side-by-side evaluation, despite lower rubric coverage
(Table~\ref{tab:official-coverage}; Figure~\ref{fig:official-preference}). Its
estimated generation cost was \$0.524 per topic, compared with \$2.823 for
\texttt{sol-research-first} (Table~\ref{tab:submissions}). However, it trailed
the leading proprietary runs on Batch-2 nugget coverage, and diagnostics
identified synthesis and revision weaknesses, including the review-pass failure
in Figure~\ref{fig:official-per-topic}. Its consistently lower Batch-1
Flesch--Kincaid scores suggest lower reading burden as a possible contributor to
human preference, although our analysis did not establish such an effect
(Section~\ref{sec:discussion}).

\textbf{RQ2.} Writer choice had the largest effect among the measured
development ablations, while the benefit of adjacent-chunk expansion varied by
writer (Section~\ref{sec:ablations}). Hybrid RRF provided the clearest
efficiency improvement: \texttt{brief-hybrid} used 14.4\% fewer searches, 8.2\%
fewer processed tokens, and had 8.5\% lower estimated generation cost than
\texttt{brief-base}. It also tied the global Batch-1 \texttt{vital\_strict}
maximum and achieved our highest \texttt{all\_strict} score in all three
official collections (Tables~\ref{tab:submissions}
and~\ref{tab:official-coverage}), although the leading proprietary systems were
not statistically distinguishable on official coverage.

\textbf{RQ3.} More elaborate orchestration did not provide a consistent
advantage across the systems we tested. \texttt{sol-research-first} improved
over \texttt{sol-default} in our development evaluation and achieved higher
Batch-1 coverage, but processed 81\% more tokens with little difference in
Batch-2 coverage. \texttt{facetrag}, despite its multi-stage architecture and
high retrieval volume, remained the weakest system on official coverage. Because
these systems differ in models, retrieval, and prompts as well as orchestration,
these comparisons do not isolate the effect of orchestration alone.

Overall, our results show that there is no single best design for agentic RAG:
different architectures offer different trade-offs in coverage, human
preference, robustness, and efficiency. The same applies to evaluation. Our
internal automated judgments, official rubric and nugget coverage, and human
side-by-side preferences sometimes produced different system rankings,
emphasizing that the choice of evaluation protocol is itself consequential.
Developing a strong system therefore requires not only selecting appropriate
models, retrieval, and orchestration, but also selecting evaluation methods that
reflect the intended qualities and use case of the system. Rather than relying
on a single metric or evaluation paradigm, future work should combine
complementary signals and validate that the chosen evaluation meaningfully
captures the behavior it is intended to measure.

\begin{acks}
  We thank the TREC 2026 RAG track organizers for their efforts in running the
  track. This work was supported by the ARC Centre of Excellence for Automated
  Decision-Making and Society (ADM+S, CE200100005) and was undertaken with the
  assistance of computing resources from RACE (RMIT AWS Cloud Supercomputing).
\end{acks}

\bibliographystyle{ACM-Reference-Format}
\bibliography{references}

\end{document}